\documentclass[aps,amssymb,amsmath,12pt,floatfix]{revtex4}

\usepackage{color}
\usepackage{subfigure}
\usepackage{graphics}
\usepackage{amssymb,amsthm,amsmath,amsfonts,latexsym}
\usepackage{rotating}
\usepackage{algorithmic}
\usepackage[sort&compress]{natbib}

\newcommand{\algorithmicgpu}{\textbf{Beginning of the GPU code}}
\newcommand{\GPU}{\item[\algorithmicgpu]}
\newcommand{\algorithmicendgpu}{\textbf{End of the GPU code}}
\newcommand{\ENDGPU}{\item[\algorithmicendgpu]}

\begin{document}

\title{Langevin dynamics simulations of biomolecules\\
on graphics processors}

\author{A. Zhmurov$^{1,2}$, R. I. Dima $^3$, Y. Kholodov$^1$ and V. Barsegov$^{2,1}$}
\thanks{Corresponding author; phone: 978-934-3661; fax: 978-934-3013;
Valeri\_Barsegov@uml.edu}
\affiliation{$^1$Moscow Institute of Physics and Technology, Dolgoprudnyi, Moscow region, Russia, 141700,\\
$^2$Department of Chemistry, University of Massachusetts, Lowell, MA 01854, \\
$^3$ Department of Chemistry, University of Cincinnati, Cincinnati, OH 45221}

\date{\today}


\begin{abstract}
Due to the very long timescales involved ($\mu s$$-$$s$), theoretical
modeling of fundamental biological processes including folding,
misfolding, and mechanical unraveling of biomolecules, under
physiologically relevant conditions, is challenging even for
distributed computing systems. Graphics Processing Units (GPUs) are
emerging as an alternative programming platform to the more
traditional CPUs as they provide high raw computational power that can
be utilized in a wide range of scientific applications. Using a
coarse-grained Self Organized Polymer (SOP) model, we have developed
and tested the GPU-based implementation of Langevin simulations for
proteins (SOP-GPU program). Simultaneous calculation of forces for all
particles is implemented using either the particle based or the
interacting pair based parallelization, which leads to a
$\sim$$30$-fold acceleration compared to an optimized CPU version of
the program. We assess the computational performance of an end-to-end
application of the SOP-GPU program, where all steps of the algorithm
are running on the GPU, by profiling the associated simulation time
and memory usage for a number of small proteins, long protein fibers,
and large-size protein assemblies. The SOP-GPU package can now be used
in the theoretical exploration of the mechanical properties of
large-size protein systems to generate the force-extension and
force-indentation profiles under the experimental conditions of force
application, and to relate the results of single-molecule experiments
{\em in vitro} and {\em in silico}.

\end{abstract}

\maketitle


\newpage

\section{Introduction}

Mechanical functions of protein fibers such as fibronectin, fibrin
fibers, microtubules, and actin filaments, are important in
cytoskeletal support and cell motility
\cite{Stossel01,DischerScience07,SchwarzBiophysJ08}, in cell adhesion
and the formation of the extracellular matrix
\cite{Leckband04,McEver02,Marshall03,Valeri05}, and in blood clotting
\cite{Weisel04,Weisel08,Lord07}. Physical properties of viral capsids
of plant and animal viruses
\cite{BrooksBJ97,WuitePNAS08,McPhersonJMB07},
retroviruses \cite{KayBJ07}, and bacteriophages
\cite{MacKintoshPNAS04,EvilevitchPNAS07}, and the transitions between
their stable and unstable states determine the life cycle of many
viruses, including virus maturation, and infection of cells
\cite{ConwayCOSB05}. Understanding the microscopic origin of the
unique viscoelastic properties of protein fibers and the crossover
from an elastic to a plastic behavior in viral capsids, as well as the
control of their mechanical response to an applied mechanical force
constitute major areas of research in biochemistry and
biophysics. Single-molecule techniques, such as AFM and laser
tweezer-based force spectroscopy, have been used to study the
mechanical properties of protein fibers
\cite{FernandezNSB03,RiefNM02,FernandezNP06,WeiselBJ07} and viral
capsids
\cite{MacKintoshPNAS04,EvilevitchPNAS07,BustamanteNature01,WuiteCellMolLifeSci07}.
Yet, due to the high complexity of these systems
($\sim$$10^3$$-$$10^5$ residues) and to their large size
($\sim50-200nm$), these experiments yield results that are nearly
impossible to interpret without first having some {\it a priori}
information about their energy landscape \cite{Weisel08}.

Standard packages for all-atom Molecular Dynamics (MD)
simulations, such as CHARMM \cite{Charmm83}, NAMD \cite{Namd05}, and
Gromacs \cite{Gromacs95} among others, are being used to access the
submolecular behaviour of biomolecules. However, because all-atomic
modeling is currently limited to a $10$$-$$50 nm$ length scale and
$0.1$$-$$10\mu s$ duration
\cite{Schulten01,Schulten07,SchultenBiophysJ08}, these methods allow
for the theoretical exploration of equilibrium properties of
biomolecules, and reaching the biologically important $ms$$-$$s$
timescale becomes virtually impossible even for a small system. More
importantly, to fully explore the free energy landscape underlying a
biological process of interest, one needs to generate a statistically
representative set of trajectories. One possibility is to carry out MD
simulations on manycore computer clusters, but it requires tremendous
computational resources and long CPU times. For example, it takes
$800,000$ CPU hours to obtain $20$ short $1ns$ MD trajectories for the
southern bean mosaic virus, which contains as many as $4.5$ million
atoms, on an SGI Altix 4700 cluster \cite{Grubmueller09}. These
limitations exclude computations as an investigative tool in the study
of a range of biological problems, such as the large deformations of
protein fibers, the formation of biomolecular complexes and
aggregates, and the mechanical failure of viral capsids, for which
experimental data are already available, thereby rendering the direct
comparison of the results of experiments \textit{in vitro} and
\textit{in silico} impossible.

Although graphics processors have been originally designed for
computationally intensive graphics rendering, they have evolved over
the last few years into highly parallel, multithreaded computing
devices. Recent technological advances in the throughput-oriented
hardware architecture of GPUs with extremely high peak arithmetic
performance, which employs IEEE floating point arithmetic, have
unleashed tremendous computational power that can now be utilized in
general purpose scientific applications. Unlike mainstream processor
architecture, GPUs devote the majority of their logic units to
performing actual calculations, rather than to cache memory and flow
control. Massive multithreading, fast context switching, and high
memory bandwidth have enabled GPUs to tolerate latencies associated
with memory calls, and to run many computational cores simultaneously
(Fig.~1). Programming tools for modern GPUs include several platforms
such as ATI Stream Computing \cite{ATIStream}, NVIDIA Compute Unified
Device Architecture (CUDA) \cite{CUDAPG,CUDABP}, and Open Computing
Language (OpenCL) \cite{OpenCLSpec}. CUDA, a parallel computing
environment (a dialect of the C and C++ programming languages),
provides a high level software platform that allows a programer to
define kernels that are executed in parallel by independent
computational threads. The cost of one GeForce GTX 295 graphics card
(from NVIDIA) with $2$ GPUs ($2$$\times$$240$ processors) is just
$\sim$$\$$$500$ (i.e. $\sim$$\$$$1$ per processor), which makes GPUs a
cheap desktop-based alternative to an expensive computer cluster.

Because GPUs differ from CPUs in several fundamental ways, CPU-based
methods for molecular simulations of biomolecules cannot be easily
translated nor simply adapted to a GPU. In these methods,
particle-particle interactions are described by the same empirical
potential energy function (force field), and the dynamics of the
system in question is obtained by solving numerically the same
equations of motion for all particles. Hence, there is a direct
correspondence between the SIMD (Single Instruction Multiple Data)
architecture of a GPU at the hardware level and the numerical routines
(software) used to follow the molecular dynamics. It is then possible
to execute ``single instruction'', i.e. calculation of the potential
energy or evaluation of forces, or generation of random forces, or
integration of the equations of motion, on ``multiple data'' sets (for
all particles) at the same time over many iterations using many
Arithmetic Logic Units (ALUs) running in parallel. This makes MD
simulations a natural candidate to implementation on a GPU, but not
all algorithms are amenable to this architecture. For an algorithm to
execute efficiently on the GPU, it must be recast into a data-parallel
form with independent threads running the same instruction stream on
different data. There exist preliminary versions of standard packages
for MD simulations of proteins implemented on a GPU, such as NAMD
\cite{Schulten07,Rodrigues08,Schulten08}, Gromacs \cite{Pande09}, and
other applications \cite{Davis09,Meel08,Anderson08}. Yet, there are no
GPU-based implementations of Langevin dynamics simulations that we are
aware of. In this paper, we develop and test such an implementation.
Because the topology and the overall structure (geometry), rather than
the atomic details, govern the force-driven molecular transitions in
protein systems, we employ a coarse-grained description of proteins
\cite{Tozzini05,ClementiJMB00,Dmitri97} using a Self Organized Polymer
(SOP) model \cite{Ruxandra06,HyeonPNAS07}.

Implementing the Langevin dynamics algorithm on a GPU requires
detailed understanding of the device architecture. In the next
Section, we review key architectural traits of modern GPUs. The
methodology for the GPU-based implementation of Langevin simulations
is presented in Section 3, where we describe the particle based and
the interacting pair based parallelization approaches to force
computation as well as the numerical routines for generating
pseudorandom numbers, constructing the Verlet lists, and integrating
forward Langevin equations to the next time step. For purposes of
presentation and to focus on the most essential computational aspects,
we have simplified the prosentation of the formalism as much as possible. A
comparative analysis of the results of CPU- and GPU-based simulations
of the mechanical unfolding for a test system represented by the
all-$\beta$-strand $WW$ domain is performed in Section 4, where we
also assess the accuracy of the numerical integration. We discuss 
the results of the GPU-based computations in terms of the
simulation time, memory usage, and computational speedup (CPU time
versus GPU time) for a range of proteins including small proteins,
such as the $WW$ domain, the $Ig27$ domain from human titin, the $C2A$
domain from human synaptotagmin ($Syt1$), the $\gamma$$C$ chain and
the double-$D$ fragment of human fibrinogen ($Fb$), single-chain
models of fibrin fibers ($Fb$ monomer and dimer), and large-size
protein assembly (viral capsid $HK97$). The main results are
summarized in Section 5.


\section{NVIDIA GPU Architecture}

The generally used CPUs have most of their logical elements dedicated
to cache and flow control to employ complex computational logic and to
provide computational cores with fast memory access (Fig.~1). This
makes a CPU capable of performing computations following a sequential
workflow. On a GPU, a large number of logical elements are devoted to
actual computations, and cache and flow control are reduced to a small
unit.  These features enable GPUs to achieve high arithmetic unit
density and to perform the same computational procedure(s)
simultaneously for all particles in a system by using many independent
threads of execution to run parallel calculations on different data
sets (Fig.~1) \cite{CUDAPG,CUDABP}. For example, on a CPU, vector
addition in $M$ dimensions is performed in a loop, where the
components of the resulting vector are computed one after another. On
a GPU, this procedure can be performed using $M$ independent threads,
each of which computes just one component. Hence, on a GPU, a vector
sum is computed at a cost of one addition, whereas on a CPU this cost
is multiplied by $M$. Yet, unlike computer clusters, where each core
is capable of following its own computational protocol, contemporary
compute-oriented GPUs are based on the SIMD architecture, which
mandates that an identical instruction stream be fed to the large
array of processing units \cite{CUDAPG}. Hence, to achieve top
performance on the GPU, one has to organize a computational task into a
data-parallel form with many independent threads performing the same
operation(s), but on different data. In addition, the task should be
compute-intensive so that, most of the time, the GPU performs actual
computations rather than reading and writing data \cite{CUDABP}.

A GPU contains several multiprocessors, each having its own flow
control and cache unit. Modern GPUs from NVIDIA (Tesla C1060, GeForce
GTX 285) host as many as $30$ multiprocessors, each with $8$ ALUs
(Fig.~1) \cite{CUDAPG}, which is a total of $240$ computational units
per chip. The GPU sits on an expansion board with its own (global or
device) memory - $512 MB$ on the more conventional GeForce cards and
up to $4 GB$ on Tesla cards - separate from the DRAM accessible by the
CPU. Although the GPU memory systems provide bandwidths of over $100
GB/s$, GPU computational cores have a low amount of cache memory,
which necessitates the use of a coalesced global memory read. Because
all the $240$ ALUs access the global memory simultaneously, the number
of memory invocations (per ALU) should be minimal to optimize the GPU
performance. For example, on graphics cards from NVIDIA, the size of
the cache memory is $\sim$$20 KB$ per multiprocessor, which is
sufficient to store the local data. However, when an ALU attempts to
access a large window of addresses stored in an on-board memory, the
cache memory may run low causing computational cores to spend more
time waiting for data (latency), rather than performing
computations. With a sufficient number of threads multiplexed on each
ALU, latency is effectively hidden and the GPU device achieves a high
performance level without the need for a cache. For example, $30,000$
threads of execution may run concurrently on the Tesla C1060 graphics
card with a peak arithmetic rate of about $900$ single precision
gigaflops. Of the devices currently available, graphics processors
produced by NVIDIA offer the most advanced as well as the most
user-friendly environment. For this reason, we used the graphics card
GeForce GTX 295 (NVIDIA), which has two GPUs, each with $240$ $1.24GHz$
ALUs ($30$ multiprocessors) and $768 MB$ of memory. Each
multiprocessor has $16KB$ of cache, of which $8KB$ is the constant
memory cache and $8KB$ is the texture memory cache (these numbers vary
with GPU make and model).

Unlike programming on a CPU, \textit{data driven} programming on a GPU
entails finding a way to partition the computational problem into many
identical subroutines, to improve the memory access pattern, and then
to optimize the numerical procedures themselves. At a software level,
a programmer can define scalar program fragments, called
\textit{kernels}, that are executed concurrently in a \textit{thread
  block} in parallel \textit{warps} of $32$ threads. A thread, an
elementary unit of the computational workflow identified by a thread
index, uses data to execute instructions that are listed in a source
code, and saves the result(s) to a unique location to avoid memory
conflicts. In a single multiprocessor, $32$$-$$512$ threads can be
combined into a thread block.  Each thread block executes a stream of
instructions within a single multiprocessor and many independent
blocks may be run concurrently across the entire GPU device
in a \textit{grid}. The software model of the GPU also includes
\textit{global}, \textit{shared}, and \textit{local} memory
\cite{CUDAPG}. Threads use their own local memory to store temporary
variables, and threads in a block can communicate through the fast
shared memory in the same multiprocessor, but threads in different
blocks cannot communicate without using the global memory, which is
particularly costly in terms of clock cycles required. Although
threads in the same block can be synchronized, there is no means of
synchronization of threads across several blocks. A particular global
memory region (software), which corresponds to the GPU on-card memory
(hardware), can be accessed by all the threads running on a GPU and by
a CPU, but global memory calls should be cached or combined due to
latency. Minimizing the number of memory calls can be achieved by
using the shared memory, data exchange among threads in one thread
block, and an optimal alignment of the computational threads with the
data arrays in the global memory (\textit{coalescent} memory access)
\cite{CUDAPG,CUDABP}. Caching the global memory can be realized by
employing \textit{texture references} or \textit{constant memory}.


\section{Langevin Dynamics Simulations on a GPU}

In this Section, we describe the particle based and the interacting
pair based methods for the parallel computation of potentials and
forces due to binary particle-particle interactions. We also outline
the numerical procedures involved in the generation of the Verlet
lists and the random forces, and in the numerical integration of the
Langevin equations of motion. We designed the algorithm to process as
many computational fragments simultaneously on a GPU as possible in
order to move the GPU into full production efficiency and to minimize
the GPU/CPU communication. The algorithm decomposition, along with the
workload division between the GPU and the CPU, is diagrammed in
Fig.~2, where we also summarize the computational workflow on a CPU
and on a GPU, including the operation on the data files for the
molecular topology, the particle energies and coordinates, and the
data flow between the CPU DRAM and the GPU global memory (host-GPU
data transfers).

In Langevin simulations of biomolecules, molecular forces are usually described by
two-body (pair) potentials, such as the harmonic potential, the FENE
potential \cite{GromacsManual}, the Lennard-Jones potential, etc.,
which differ in their mathematical form. Hence, the same generic
algorithm can be employed to compute these potentials. In the
pseudocode listings below, the $i$-th residue for a protein of $N$
residues has a unique index $i\in[0, N-1]$; for an array of particle
coordinates $\vec{r}$, $\vec{r}[i]$ denotes coordinates of the $i$-th
particle. We also use the following notations: global memory reads are
represented by $\Leftarrow$, saving data to the GPU global memory is
shown by $\Rightarrow$, $\leftleftarrows$ denotes cached texture
memory reads, and $\leftarrow$ represents local, shared, or constant
memory invocation, or assignments of variables. Consider the following
computational procedure for evaluating forces on a CPU:

\noindent
{\bf Algorithm 1:} Calculation of pairwise forces (CPU implementation).

\algsetup{indent=2em,linenodelimiter=.}
\begin{algorithmic}[1]
\STATE $P \leftarrow $ total number of pairs in a system for a given potential
\FOR {$p=0$ to $P-1$}
\STATE $i \leftarrow pairs[p].i$ \COMMENT{index of the first particle in a particle pair}
\STATE $j \leftarrow pairs[p].j$ \COMMENT{index of the second particle in a particle pair}
\STATE $par \leftarrow pairs[p].parameters$ \COMMENT{parameters for the $i$$-$$j$ pair}
\STATE $\vec{r} \leftarrow \vec{r}[i]-\vec{r}[j]$
\STATE $\vec{df} \leftarrow \texttt{force(}\vec{r}, par\texttt{)}$
\STATE $\vec{f}[i] \leftarrow \vec{f}[i] \pm \vec{df}$
\STATE $\vec{f}[j] \leftarrow \vec{f}[j] \mp \vec{df}$
\ENDFOR
\end{algorithmic}

Information about the residue pairs is stored in array $pairs$, which
has indices $i$ and $j$, and variable $par$ for constant parameters,
which specify the potential energy. In the main cycle on a CPU (line
$2$), index $p$ runs through all the pairs of interacting residues
$p$$=$$0,1,\ldots,P-1$. For each pair, the information about the
coordinates ($\vec{r}[i]$ and $\vec{r}[j]$) and the constant
parameters ($par$) is gathered in lines $3$$-$$5$. The force increment
$\vec{df}$ is computed in $\texttt{force(\ldots)}$ for a given pair
$p$ (line $7$). This value is added (or subtracted) to (from) the
force for the $i$-th particle ($\vec{f}[i]$, line $8$) and subtracted
(or added) from (to) the force for the $j$-th ($\vec{f}[j]$, line
$9$). On a CPU, the force values are computed only once (lines
$7$$-$$9$). Because force calculations are sequential, they do not
overlap in time. By contrast, because on a GPU a two-body potential is
computed for different pairs of residues in different threads, a naive
addition (lines $8$ and $9$) may cause memory conflicts when some or
all the threads attemp to access the same address in the GPU global
memory.

There are two main optimization strategies that allow one to avoid
this situation. In the first approach, all the forces for one particle
are computed in one thread, which requires running $N$ threads to
obtain the force values for all particles. We refer to this procedure
as the particle based parallelization approach. The use of this
approach results in the same force, acting on the $i$-th and $j$-th
particles, but computed twice in the $i$-th and $j$-th threads
\cite{Anderson08}. Following a different strategy, which we refer to
as the interacting pair based parallelization approach, force
calculations are performed for all pairs in parallel using $P$
independent theads, and $2$$P$ force values are saved to different
locations in the GPU global memory. We pursued in detail both
optimization strategies, which exploit the data-parallel aspects of
GPU based computing.


\subsection{The particle based parallelization approach}

In this approach, $N$ independent threads run on a GPU concurrently,
each computing all the pair potentials for each particle and summing
all force values (except for the random force) to obtain the total
force. Although the force acting on the $i$-th and $j$-th particles is
computed twice, the number of global memory calls is reduced by a
factor of $2$$N$, and the time spent on recalculating the same
potential is compensated by the time saved by not waiting to write the
force data to and to read the data from the GPU global memory.

\noindent
{\bf Algorithm 2:} Calculation of pairwise forces using particle based parallelization.

\begin{algorithmic}[1]
\STATE $\vec{f}_i \leftarrow 0$ \COMMENT{resulting force}
\STATE $i \leftarrow $ GPU thread index \COMMENT{same as particle index}
\STATE $\vec{r}_i \leftleftarrows \vec{r}[i]$ \COMMENT{coordinates of the $i$-th particle}
\STATE $P_i \Leftarrow P_p[i]$  \COMMENT{number of pairs formed by the $i$-th particle}
\FOR {$p=0$ to $P_i-1$}
\STATE $j \Leftarrow PairsMap[i][p].j$ \COMMENT{second particle in a pair}
\STATE $\vec{r}_j \leftleftarrows \vec{r}[j]$ \COMMENT{coordinates of the $j$-th particle}
\STATE $par \Leftarrow PairsMap[i][j].parameters$ \COMMENT{parameters for the $i$$-$$j$ pair}
\STATE $\vec{r} \leftarrow \vec{r}_i-\vec{r}_j$
\STATE $\vec{df} \leftarrow \texttt{force(}\vec{r},par\texttt{)}$
\STATE $\vec{f}_i \leftarrow \vec{f}_i \pm \vec{df}$
\ENDFOR
\STATE Output: $\vec{f}_i$
\end{algorithmic}

The array $P_p$ keeping track of the number of pairs for all residues
and the matrix $PairsMap$ of all particle pairs are pre-generated on a
CPU and fetched to the GPU global memory. $P_p$ is an $N$-dimensional
vector of integers $0,1,2,\ldots$.  Each element of this vector
corresponds to a single particle, and the $i$-th integer value is the
number of particles interacting with the $i$-th particle. $PairsMap$
is the $N$$\times$$M$ matrix, where $M$ is a maximum number from array
$P_p$. The $i$-th row of the matrix $PairsMap$ corresponds to the
$i$-th particle and contains the indices of all particles interacting
with the $i$-th particle and the constant parameters for the potential
energy function.  Data in the $PairsMap$ can be easily re-arranged for
coalescent memory reads. Coordinates of the second particle for each
pair are accessed at random, which allows one to take advantage of the
texture reference in global memory reads. Force values are computed in
parallel by $N$ threads as follows. First, the $i$-th thread reads the
coordinates $\vec{r}_i$ using a cached texture reference (line $3$)
and the number of particle pairs $P_p[i]$ (line $4$). By cycling
through all the pairs, formed by the $i$ residue (lines $5$$-$$12$),
the thread reads the index $j$ and the coordinates $\vec{r}_j$ of the
$j$-th particle and the constant parameters ($par$, lines
$6$$-$$8$). These are used to compute the force increment
$\vec{df}$ (line $10$), which is added to the resulting force
$\vec{f}$ (line $11$) for the $i$-th particle. The parameters $par$
depend on the potential used. For example, for a covalent bond,
described by a harmonic potential
$V_H$$=$$K^{sp}_{ij}(r_{ij}-r^0_{ij})^2/2$, $par$ contains the
equilibrium distance $r^0_{ij}$ and the spring constant $K^{sp}_{ij}$.

\noindent
{\bf Verlet lists:} In molecular simulations, the information about the covalent bonds and the native interactions (array 
$P_p$ and matrix $PairsMap$), obtained from the PDB structure of a protein, does not change.  However, the 
information about nonbonded long-range interactions, describing the gradual attraction and hardcore repulsion between
pairs of atoms, needs to be updated from time to time. This is the most computationally demanding component of the
algorithm, since the complexity of the calculation is $O(N^2)$. A common approach is to take advantage of the fact that 
long-range interactions vanish over some distance. This allows one to use pair lists that include pairs 
of particles that are closer than the cutoff distance (Verlet lists) \cite{Verlet73}. In the particle based parallelization 
approach, the array $P_p$ and the matrix $PairsMap$ have to be regenerated on a GPU in order to accelerate the computation 
of the potential energy using Verlet lists. This can be done by rearranging the pseudocode for particle based parallelization 
(Algorithm 2):

\noindent
{\bf Algorithm 3:} Calculation of forces using particle based parallelization and Verlet lists.

\begin{algorithmic}[1]
\STATE $i \leftarrow $ GPU thread index \COMMENT{same as particle index}
\STATE $p_i \leftarrow 0$ \COMMENT {counter of residue pairs in Verlet list}
\STATE $\vec{r_i} \leftleftarrows \vec{r}[i]$ \COMMENT{coordinates of the $i$-th particle}
\STATE $P_{p,i} \Leftarrow P_{pp}[p_1]$  \COMMENT{number of all pairs for the $i$-th particle}
\FOR {$p_p=0$ to $P_{p,i}-1$}
\STATE $j \Leftarrow PossiblePairsMap[i][p_p]$ \COMMENT{second $j$-th particle in a pair}
\STATE $\vec{r_j} \leftleftarrows \vec{r}[j]$ \COMMENT{coordinates of the $j$-th particle}
\STATE $r \leftarrow \vert\vec{r}[i]-\vec{r}[j]\vert$
\IF {$r < cutoff$}
\STATE $PossiblePairsMap[i][p_p] \Rightarrow PairsMap[i][p_i]$
\STATE $p_i \leftarrow p_i+1$
\ENDIF
\ENDFOR
\STATE $p_i \Rightarrow P_p[i]$
\end{algorithmic}

Cycling over $p_p$ includes all possible residue pairs to identify
pairs that are within the cutoff distance. The interparticle
distances are computed in line $8$. The number of particles $p_i$ that
are within the cutoff distance is counted (line $11$), and a newly
found pair is added to the matrix $PairsMap$ at the
$(i,p_i)$-position, i.e. copied from the matrix $PossiblePairsMap$
(the map of all pairs) to the matrix $PairsMap$ (the map of
pairs). Once the cycle is completed, $p_i$ is saved to the array
$P_p[i]$, which stores the numbers of the residue pairs in the Verlet
list.


\subsection{The interacting pair based parallelization approach}

To avoid computing the two-body potentials on a GPU twice, one can
design a different computational algorithm, where each thread
calculates a single pair potential for two coupled residues.  Then,
forces acting on the interacting particles in opposite directions are
computed only once, the force values obtained are saved to different
locations in the GPU global memory, and all the forces exerted on each
particle are summed up to obtain the total force. This approach
requires additional memory calls and a gathering subroutine for the
force summation, but it enables one to accelerate simulations when the
number of residues $N$ is of the same order of magnitude as the number
of ALUs, and/or when the computation of pair potentials is
expensive. In the following pseudocode for the force calculation, $P$
(number of threads) is equal to the number of interacting pairs for
just one potential energy term, and each thread computes forces for
one pair of residues:

\noindent
{\bf Algorithm 4:} Calculation of pairwise forces using interacting pair based parallelization.

\begin{algorithmic}[1]
\STATE $p \leftarrow $ GPU thread index \COMMENT{same as pair index}
\STATE $pair \Leftarrow pairs[p]$
\STATE $par \Leftarrow PairsParameters[p]$ \COMMENT{parameters for one pair of residues $p$}
\STATE $i \leftarrow pair.i$ \COMMENT{the $i$-th particle in the pair}
\STATE $j \leftarrow pair.j$ \COMMENT{the $j$-th particle in the pair}
\STATE $shift_i \leftarrow pair.shift_i$ \COMMENT{position in array of forces for the $i$-th particle}
\STATE $shift_j \leftarrow pair.shift_j$ \COMMENT{position in array of forces for the $j$-th particle}
\STATE $\vec{r}_i \leftleftarrows \vec{r}[i]$ \COMMENT{coordinates of the $i$-th particle}
\STATE $\vec{r}_j \leftleftarrows \vec{r}[j]$ \COMMENT{coordinates of the $j$-th particle}
\STATE $\vec{r} \leftarrow \vec{r}_i-\vec{r}_j$
\STATE $\vec{df} \leftarrow \texttt{force(}\vec{r},par\texttt{)}$
\STATE $\pm\vec{df} \Rightarrow \vec{F}[i][shift_i]$ \COMMENT{saving force for the $i$-th particle}
\STATE $\mp\vec{df} \Rightarrow \vec{F}[j][shift_j]$ \COMMENT{saving force for the $j$-th particle}
\end{algorithmic}

Each thread identifies one pair potential using the thread index $p$,
and reads the information about the potential from vectors $pairs$ and
$PairsParameters$ about the constant parameters ($par$), the particle
identity ($i$ and $j$) and particle coordinates ($\vec{r}_i$ and
$\vec{r}_j$), and the global memory addresses for saving the force
values ($shift_i$ and $shift_j$). In the array $F$, the force values
are saved to the position defined by the particle index $i$ or $j$
($i$-th or $j$-th row) and by the parameters $shift_i$ or $shift_j$
(columns). In the array $pairs$, each output position $i$ and
$shift_i$ for the $i$-th particle has to be unique so that each force
value computed is saved to a different address in the GPU global
memory. This allows one to avoid memory conflicts, but requires an
additional gathering kernel for summing all the forces for a given
particle obtained in the array $\vec{F}$ (in Algorithm 4):

\noindent
{\bf Algorithm 5:} Gathering kernel for the force computation.

\begin{algorithmic}[1]
\STATE $i \leftarrow $ GPU thread index \COMMENT{same as particle index}
\STATE $\vec{f}_i \leftarrow 0$ \COMMENT{resulting force due to one potential energy term}
\STATE $P_{i} \Leftarrow P_p[i]$  \COMMENT{number of particle pairs for the $i$-th particle}
\FOR {$p=0$ to $P_{i}-1$}
\STATE $\vec{df} \Leftarrow \vec{F}[i][p]$
\STATE $\vec{f}_i \leftarrow \vec{f}_i + \vec{df}$
\ENDFOR
\STATE Output: $\vec{f}_i$
\end{algorithmic}

$P_i$ counts the residue pairs for the $i$-th particle (for one
potential energy term). The total force for the $i$-th residue
($\vec{f}_i$) is calculated by summing over all the forces computed
previously ($\vec{F}[i][p]$, line $12$$-$$13$). This part of the
program can be incorporated into the integration kernel to minimize
the number of computational kernels. On GPUs with the new Fermi
architecture (from NVIDIA), the use of thread safe atomic addition of
the computed force values to a specific location in the GPU global
memory will help to remove the performance barriers associated with
multiple memory calls.

\noindent
{\bf Verlet lists:} In the interacting pair based parallelization
approach, generating a Verlet list surmounts to forming the vector
$pairs$ of all residue pairs for one potential energy term. On a GPU,
constructing this vector is a formidable task, since the exact
position in the list, to which information about the next residue pair
should be saved, is not known. One possibility is to use the
\texttt{atomicAdd(...)} routine from the CUDA Software Development Kit
\cite{CUDAPG}, which allows one to add integers in the GPU global
memory without running into memory conflicts even when many threads
attempt to access the same memory address at the same time. However,
when many threads run in parallel, identifying new pairs and saving
them one after another may result in a Verlet list that is not ordered
according to the particle index. This may result, in turn, in an
inefficient utilization of the cache memory. To obtain an ordered
Verlet list, it has to be sorted or updated on a CPU. It is more
efficient to compute interparticle distances on a GPU, copy them to
the CPU DRAM, and then generate a new list.


\subsection{The random force}

Langevin simulations require a reliable source of $3$$N$ normally
distributed pseudorandom numbers, $g_{i,\alpha}$ ($\alpha$$=$$x$, $y$,
and $z$) produced at each integration step, in order to compute the
three components of a Gaussian random force
$G_{i,\alpha}$$=$$g_{i,\alpha}$$\sqrt{2k_BT\xi h}$, where $\xi$ is the
friction coefficient, and $h$ is the integration time step. A
pseudorandom number generator (PRNG) produces a sequence of random
numbers $u_{i,\alpha}$ uniformly distributed in the unit interval
$[0,1]$. This sequence, which imitates a sequence of independent and
identically distributed (i.i.d.) random variables, is then translated
into the sequence of normally distributed pseudorandom numbers with
zero mean and unit variance ($g_{i,\alpha}$) using the Box-Mueller
transformation \cite{Box58}. While there exist stand alone
implementations of good quality PRNGs on a GPU, in Langevin
simulations a PRNG should be incorporated into the integration kernel
to minimize read/write calls of the GPU global memory.

The simplest approach for constructing a PRNG on a GPU is to initiate
an independent generator in each thread (one-PRNG-per-thread approach)
so that pseudorandom numbers can be produced during the numerical
integration of the Langevin equations. First, a CPU generates $N$
independent sets of random seeds for $N$ PRNGs, and then transfers
them to the GPU global memory. When $4$$N$ i.i.d. pseudorandom numbers
$u_{i,\alpha}$ are needed for generating $3$$N$ normally distributed
random numbers $g_{i,\alpha}$, each thread reads a corresponding set
of random seeds to produce $4$ normally distributed random numbers for
each residue. Then, a PRNG updates its current state in the GPU global
memory, which is used as an initial seed within the same thread at the
next time step. In a different appproach, random seeds for just one
PRNG state can be shared among the computational threads on the entire
GPU (one-PRNG-for-all-threads approach). Using both approaches, we
have developed and tested GPU-based realizations of PRNG which are
based on the Hybrid Taus, Ran$2$, Lagged Fibonacci and Mersenne Twister
algorithms (manuscript in preparation). It has been shown that these
algorithms pass a number of stringent statistical tests and produce
pseudorandom numbers of very high statistical quality \cite{GPUGems3}.

\noindent
{\bf{Hybrid Taus algorithm:}} In this paper, we have implemented on a
GPU the Hybrid Taus generator \cite{GPUGems3} - Tausworthe algorithm
combined with a Linear Congruential Generator (LCG) - employing the
one-PRNG-per-thread approach.  Hybrid Taus PRNG uses a small memory
area since only four integers are needed to store its current
state. Tausworthe taus$88$ is a fast equi-distributed modulo $2$
generator \cite{Tausworthe65,Lecuyer96}, which produces pseudorandom
numbers by generating a sequence of bits from a linear recurrence
modulo $2$, and forming the resulting number by taking a block of
successive bits. In the space of binary vectors, the $i$-th element of
a vector is constructed using the following transformation:
$y_i$$=$$a_1$$y_{i-1}$$+$$a_2$$y_{i-2}$$+$$\ldots$$a_k$$y_{i-k}$,
where $a_i$ are constant coeffitients. Given the initial values,
$y_0,y_1,\ldots y_{i-1}$, the $i$-th random integer is obtained as
$x_i$$=$$\sum_{j=1}^L$$y_{is+j-1}$$2^{-j}$, where $s$ is a positive
integer and $L$$=$$32$ is the machine word size. When
$a_k$$=$$a_q$$=$$a_0$$=$$1$, where $0$$<$$2q$$<k$, and $a_i$$=$$0$ for
$0$$<$$s$$\leq$$k$$-$$q$$<$$k$$\leq$$L$, the algorithm can be
simplified to a series of binary operations
\cite{Lecuyer96}. Statistical qualities of pseudorandom numbers
produced using the taus$88$ algorithm are high \cite{GPUGems3} when
taus$88$ is combined with the LCG. When the periods of several components
of a generator are co-prime numbers, the period of a combined
generator is the product of the periods of all the components. A
similar approach is used in the KISS algorithm, which combines the
LCG, Tausworthe and two multiple-with-carry generators
\cite{Marsaglia99}. We used constant parameters which resulted in
periods of $2^{31}-1$, $2^{30}-1$, and $2^{28}-1$ for three Tausworthe
generators and a period of $2^{32}$ for the LCG generator; the period
of the combined generator is
$\sim$$2^{121}$$>$$2$$\times$$10^{36}$. In the pseudocode below,
superscripts $h$ and $d$ denote the host (CPU) and the device (GPU)
memory, respectively; a section of the code, executed on the GPU, is
in the same listing as the code for the CPU. In the CUDA program, the
corresponding code for the GPU is organized into a separate kernel:

\noindent
{\bf Algorithm 6:} Hybrid Taus PRNG.

\algsetup{
	indent=2em,
	linenodelimiter=.
}
\begin{algorithmic}[1]
\REQUIRE $y_1^h[N]$, $y_2^h[N]$, $y_3^h[N]$ and $y_4^h[N]$ allocated in CPU memory
\REQUIRE $y_1^d[N]$, $y_2^d[N]$, $y_3^d[N]$ and $y_4^d[N]$ allocated in GPU global memory
\STATE $y_1^h[1\ldots N]$ to $y_4^h[1\ldots N] \leftarrow$ Initial seeds. 
\STATE $y_1^h[1\ldots N]$ to $y_4^h[1\ldots N] \rightarrow y_1^d[1\ldots N]$ to $y_4^d[1\ldots N]$ 
\COMMENT{copying initial seeds to GPU}
\GPU{}
\STATE $j_{th} \leftarrow$ thread index
\STATE $y_1$, $y_2$, $y_3$ and $y_4 \Leftarrow y_1^d[j_{th}]$, $y_2^d[j_{th}]$, $y_3^d[j_{th}]$ and $y_4^d[j_{th}]$ 
\COMMENT{loading the state of a generator}
\FOR[four pseudorandom numbers are to be generated]{$i=1$ to $4$} 
\STATE $b \leftarrow (((y_1 \ll c_{11}) \;\mathrm{XOR}\;y_1) \gg c_{21})$
\STATE $y_1 \leftarrow (((y_1 \;\mathrm{AND}\; c_1) \ll c_{31})\;\mathrm{XOR}\;b$
\STATE $b \leftarrow (((y_2 \ll c_{12}) \;\mathrm{XOR}\;y_2) \gg c_{22})$
\STATE $y_2 \leftarrow (((y_2 \;\mathrm{AND}\; c_2) \ll c_{32})\;\mathrm{XOR}\;b$
\STATE $b \leftarrow (((y_3 \ll c_{13}) \;\mathrm{XOR}\;y_3) \gg c_{23})$
\STATE $y_3 \leftarrow (((y_3 \;\mathrm{AND}\; c_3) \ll c_{33})\;\mathrm{XOR}\;b$
\STATE $y_4 \leftarrow ay_4+ c$
\STATE Output $mult\times(\;\mathrm{XOR}\;y_1\;\mathrm{XOR}\;y_2\;\mathrm{XOR}\;y_3\;\mathrm{XOR}\;y_4)$ 
\ENDFOR \COMMENT{generating the next random number}
\STATE $y_1$, $y_2$, $y_3$ and $y_4 \Rightarrow y_1^d[j_{th}]$, $y_2^d[j_{th}]$, $y_3^d[j_{th}]$ and $y_4^d[j_{th}]$ 
\COMMENT{saving the current state of the generator}
\ENDGPU{}
\end{algorithmic}

In this listing, $b$ is a temporary unsigned integer variable, $y_1$, $y_2$, $y_3$, and $y_4$ are unsigned integer 
random seeds for three Tausworthe generators (lines $6$ to $11$) and one LCG (line $12$), and $\mathrm{XOR}$ is a 
binary operation of exclusive disjunction; $mult$$=$$2.3283064365387$$\times$$10^{-10}$ is a multiplier that converts 
a resulting integer into a floating point number (between $0$ and $1$); $c_{11}$$=$$13$, $c_{21}$$=$$19$, $c_{31}$$=$$12$, 
$c_{21}$$=$$2$, $c_{22}$$=$$25$, $c_{23}$$=$$4$, $c_{31}$$=$$3$, $c_{32}$$=$$11$, $c_{33}$$=$$17$, $c_1$$=$$4294967294$, 
$c_2$$=$$4294967288$, and $c_3$$=$$4294967280$ are constant parameters for the Tausworthe generators \cite{Lecuyer96},
and $a$$=$$1664525$ and $c$$=$$1013904223$ are constant parameters for the LCG \cite{Recipes}.


\subsection{The numerical integration kernel}

On a GPU, the Langevin equations of motion can be solved
simultaneously for all $N$ particles in $N$ threads working in
parallel. When the particle based parallelization is utilized, the
subroutines for the force computation can be incorporated into the
integration kernel. This allows a programmer to use coordinate
variables, stored locally in the GPU global memory, that are read only
once at the beginning of the computational procedure and are passed to
the next subroutine. Since all the interactions are more or less
local, texture cache can be used as well to access the coordinates in
the GPU global memory. When the interacting pair based parallelization
is employed, the force computations can be performed in a separate
kernel and the summation of all the forces (gathering kernel,
Algorithm 5) can be done inside the integration kernel. Using one
kernel for the force computation, the force summation and the
numerical integration minimizes the number of kernel invocations on
the CPU, thus, saving time for context switching on the GPU.

\noindent
{\bf Algorithm 7:} Numerical integration of the Langevin equations of motion.

\begin{algorithmic}[1]
\STATE $i \leftarrow $ GPU thread index \COMMENT{same as particle index}
\STATE $vec{r}_i \leftleftarrows \vec{r}[i](t_n)$ \COMMENT{reading coordinate of the $i$-th 
particle at the beginning of each step}
\STATE $\vec{f}_i \leftarrow \sum_v\vec{f}_{i,v}$ \COMMENT{total force exerted on the $i$-th particle due to several pair potentials}
\STATE $\vec{g}_i \leftarrow (g_x, g_y, g_z)$ \COMMENT{$3$D vector with $3$ normally distributed random numbers}
\STATE $\vec{r}_i \leftarrow \vec{r}_i + \vec{f}_i h/\xi + \vec{g}_i \sqrt{2kTh/\xi}$
\COMMENT{$1$-st order integration scheme}
\STATE $\vec{r}_i \Rightarrow \vec{r}[i](t_{n+1})$ \COMMENT{saving coordinates to global memory at the end of each step}
\end{algorithmic}

Each thread computes the displacement vector for just one
particle. Once particle coordinates are retrieved from the GPU global
memory via texture reference (line $2$), they are used in the
computational procedures that follow (the total force is computed in
line $3$). The corresponding forces $\vec{f}_{i,v}$, where the index
$v$ is running over different potential energy terms, are computed
using ether the particle based or the interacting pair based
parallelization approach. When the former approach is used, the entire
computational procedure from the force computation to the random force
generation, and to the numerical integration can be organized into a
single kernel. In the latter case, an additional gathering kernel is
needed to compute the total force (line $3$). Continuing, particles
are shifted to their new positions (line $5$), which are saved to the
GPU global memory (line $6$). Since these coordinates are used at the
next time step $t_{n+1}$, they have to be moved from the time layer
$r[i](t_{n+1})$ to the time layer $r[i](t_n)$ at the end of each
iteration.


\section{The SOP-GPU program for Langevin simulations of proteins}

\subsection{The SOP model}

We employed the methodology for the GPU-based realization of Langevin
dynamics to develop a CUDA program for biomolecular simulations fully
implemented on a GPU. To describe the molecular force field, we
adapted the Self Organized Polymer (SOP) model (SOP-GPU program)
\cite{Ruxandra06}. Previous studies have shown that the SOP model
describes well the mechanical properties of proteins, including the
Green Fluorescent Protein \cite{Ruxandra07} and the tubulin dimer
\cite{Ruxandra08}. In the SOP model, each residue is described using a
single interaction center ($C_{\alpha}$-atom). The potential energy
function of a protein conformation $V$, specified in terms of the
coordinates $\{ r\}$$=$$r_1,r_2,\ldots,r_N$, is given by
\begin{equation}\label{eq:sop}
\begin{split}
V&=V_{FENE}+V^{ATT}_{NB}+V^{REP}_{NB}=\\
&-\sum_{i=1}^{N-1}{\frac{k}{2}R_0^2\log{\left(1-\frac{\left(r_{i,i+1}-r^0_{i,i+1}\right)^2}{R_0^2}\right)}}\\
&+\sum_{i=1}^{N-3}{\sum_{j=i+3}^{N}{\varepsilon_n\left[\left(\frac{r^{0}_{ij}}{r_{ij}}\right)^{12}-
2\left(\frac{r^{0}_{ij}}{r_{ij}}\right)^{6}\right]\Delta_{ij}}}\\
&+\sum_{i=1}^{N-2}{\varepsilon_r\left(\frac{\sigma_{i,j+2}}{r_{i,j+2}}\right)^{6}}+\sum_{i=1}^{N-3}
{\sum_{j=i+3}^{N}{\varepsilon_r\left(\frac{\sigma}{r_{ij}}\right)^{6}\left(1-\Delta_{ij}\right)}}.
\end{split}
\end{equation}
In Eq.~(\ref{eq:sop}), the finite extensible nonlinear elastic (FENE)
potential $V_{FENE}$ describes the backbone chain connectivity. The
distance between two next-neighbor residues $i$ and $i$$+$$1$, is
$r_{i,i + 1}$, while $r^0_{i,i+1}$ is its value in the native (PDB)
structure, and $R_0$$=$$2$\AA\ is the tolerance in the change of a
covalent bond (first term in Eq.~(\ref{eq:sop})). We used the
Lennard-Jones potential ($V^{ATT}_{NB}$) to account for the
non-covalent interactions that stabilize the native state (second term
in Eq.~(\ref{eq:sop})). We assumed that, if the noncovalently linked
residues $i$ and $j$ ($|i-j|$$>$$2$) are within the cutoff distance
$R_C$$=$$8$\AA, then $\Delta_{ij}$$=$$1$, and zero otherwise. We
used a uniform value for $\epsilon_n$$=$$1.5 kcal/mol$, which
quantifies the strength of the non-bonded interactions. All the
non-native interactions in the $V^{REP}_{NB}$ potential are described
as repulsive (third term in Eq.~(\ref{eq:sop})). Additional constraint
are imposed on the bond angle formed by residues $i$, $i$$+$$1$, and
$i$$+$$2$ by including the repulsive potential with parameters
$\epsilon_r$$=$$1 kcal/mol$ and $\sigma_{i,i+2}$$=$$3.8$\AA, which
determine the strength and the range of the repulsion. To ensure the
self-avoidance of the protein chain, we set $\sigma$$=$$3.8$\AA\ (last
term in Eq.~(\ref{eq:sop})).


\subsection{Benchmark simulations}

We carried out test simulations of the mechanical unfolding for the
all-$\beta$-strand domain $WW$ from the human $Pin1$ protein (PDB code
1PIN, Table I) using the SOP-GPU program. The rationale behind
choosing this protein as a test system is two-fold. First, the $WW$
domain is of particular interest to the field of protein folding and
dynamics, and several research groups have expended considerable
efforts to characterize the biophysical and biochemical properties of
this protein
\cite{FergusonPNAS01,KaranicolasPNAS03,SchultenBiophysJ08}. Secondly,
this is the smallest known independently folding all-$\beta$ domain
and the all-$\beta$ protein architecture is the primordial structural
state that can be studied experimentally using single-molecule force
spectroscopic techniques such as AFM, and laser and optical tweezers
\cite{RiefScience97,DietzPNAS04}. For these reasons, the $WW$ domain
has been extensively used in the theoretical exploration of protein
folding and unfolding.

We consider the following principal sources of error: $(1)$ precision
issues arising from the differences in single precision (GPU) and
double precision (CPU) IEEE floating point arithmetic, $(2)$ possible
read/write errors in the GPU global memory (hardware), and $(3)$
accuracy of the SOP-GPU program, i.e. possible errors in the numerical
routines (software). We report on our implementation of the SOP-GPU
package on the NVIDIA GeForce GTX 295 (Section II) and compare it
against a dual Quad Core Xeon $2.66 GHz$, considered to be
representative of similar levels of technology. All CPU/GPU benchmarks
have been obtained on a single GPU and a single CPU. To obtain the
dynamics of the force-induced molecular elongation, the Langevin
equations of motion for each residue ${\bf{r}}_i$ have been integrated
numerically using the first-order integration scheme (in powers of the
integration time step $h$) \cite{Ermak78},
\begin{equation}\label{Ermak}
{\bf{r}}_i(t+h)={\bf{r}}_i(t)+f({\bf{r}}_i(t))\xi h+G_i(t),
\end{equation}
where $G_i$ is the random force, and
$f({\bf{r}}_i)$$=$$-$$(\partial{V({\bf{r}}_i)}/\partial{{\bf{r}}_i})$
is the total force due to the covalent and the noncovalent
interactions (Eq.~(\ref{eq:sop})) exerted on the $i$-th
particle. Benchmark simulations of the mechanical unfolding of the
$WW$ domain have been carried out at room temperature
($k_B$$T$$=$$4.14 pN/nm$) over $4$$\times$$10^8$ iterations with the
time step $h$$=$$20 ps$, using the standard bulk water viscosity
($\xi$$=$$7.0\times 10^5 pN ps/nm$). Each trajectory has been
generated by fixing the $N$-terminal end and pulling the $C$-terminal
end of the $WW$ domain with the time-dependent mechanical force
$f_{ext}(t)$$=$$r_f$$t$ in the direction corresponding to the
end-to-end vector, and using the force-loading rate
$r_f$$=$$\kappa$$\nu_0$, where $\kappa$$=$$35 pN/nm$ is the cantilever
spring constant and $\nu_0$$=$$2.5 \mu m/s$ is the pulling speed.

The results of the CPU- and the GPU-based computations are presented
in Fig.~3, where we compare the force-extension curves $f(R)$ and the
average temperature $\langle T\rangle$ for two representative
trajectories of unfolding, and the distributions of unfolding forces
$p(f^*)$ sampled from $260$ trajectories on a CPU and $300$
trajectories on a GPU. Temperature conservation ($d\langle
T\rangle$$/dt$), mechanical work performed on the system
($w$$=$$\int_{R_0}^{R_{fin}}f(R)dR$), and the distribution of peak
forces ($f^*$'s) are rigorous physical metrics for measuring the
precision of molecular simulations.  Aside from small deviations due
to the different initial conditions, the profiles of $f(R)$ and
$\langle T\rangle$, obtained on the CPU, are close to the profiles of
the same quantities, generated on the GPU. A small drop in $\langle
T\rangle$ is due to the onset of the unfolding transition in the $WW$
domain, which occurs at $t$$\approx$$0.15 ms$. Both the CPU- and the
GPU-based calculations of the histogram of unfolding forces $p(f^*)$
result in similar values of the average force, i.e.  $\langle
f^*\rangle$$\approx$$120.56 pN$ on the CPU and $\langle
f^*\rangle$$\approx$$120.94 pN$ on the GPU, but slightly different
standard deviations of $\sigma_{f^*}$$\approx$$5.83 pN$ on the CPU
versus $\sigma_{f^*}$$\approx$$6.58 pN$ on the GPU due to the small
sample size (Fig.~3). The magnitude of the critical force for
unfolding is well within the $60 pN$$-$$200 pN$ force range observed
for mostly $\beta$-strand single domain proteins
\cite{RiefScience97,CarrionVazquezProgBiophysMolBiol00}.


\subsection{Accuracy of the numerical integrators}

Langevin simulations fully implemented on a GPU enable one to obtain
long trajectories of protein dynamics generated over as many as
$10^9$$-$$10^{10}$ iterations. Consequently, there emerges a question
about the numerical accuracy of the integration scheme used. In
Langevin simulations of proteins, the equations of motion are solved
numerically using the first-order integrator (Eq.~(\ref{Ermak}))
\cite{Ermak78}. However, the magnitude of the associated numerical
error, which may, potentially, add up over many billions of
iterations, is not known. We assessed the numerical accuracy of
integration protocols by considering the mechanical unfolding of a
protein. To test the results of simulations of the protein extension
$\Delta$$X(t)$$=$$X(t)$$-$$X_0$ against the theoretical predictions,
we used an exactly solvable model of a Brownian particle $X(t)$
evolving in a one-dimensional harmonic potential,
$V(X)$$=$$K_{sp}(X-X_0)^2/2$, where $X_0$ is the equilibrium position
and $K_{sp}$ is the molecular spring constant \cite{Doi}.

Under the non-equilibrium conditions of the time-dependent force application, $f_{ext}(t)$$=$$r_f$$t$, the average 
particle position (the end-to-end distance), computed theoretically, is given by
\begin{equation}\label{eq:Xav}
\langle X(t)\rangle_{th} = X_0 e^{-t/\tau} + \frac{r_f \tau}{\xi}\left( t-\tau \left( 1-e^{-t/\tau}
\right) \right),
\end{equation}
where $\tau$$=$$\xi$$/K_{sp}$ is the characteristic timescale. In pulling simulations, the average particle position at 
the step $n$$+$$1$, $\langle X(t_{n+1})\rangle$, where $t_n$$=$$n$$h$, can be obtained recursively from the average 
position obtained at the previous $n$-th step, $\langle X(t_{n})\rangle$, using the first-order integration scheme,
\begin{equation}\label{eq:1st}
\langle X(t_{n+1})\rangle_{sim} = \langle X(t_n)\rangle + \left( \frac{r_f t_n}{\xi} - 
\frac{\langle X(t_n)\rangle }{\tau}\right) h,
\end{equation}
or the second-order integration scheme,
\begin{equation}\label{eq:2nd}
\langle X(t_{n+1})\rangle_{sim} = \langle X(t_n)\rangle + \left( \frac{r_f t_n}{\xi} - 
\frac{\langle X(t_n)\rangle }{\tau}\right) h + \left( \frac{\langle X(t_n)\rangle}{2\tau^2} - 
\frac{r_f t_n}{2\tau \xi }\right) h^2
\end{equation}
Hence, Eq.~(\ref{eq:Xav}), (\ref{eq:1st}), and (\ref{eq:2nd}) can be used to assess the accuracy of the numerical 
integrators.

We carried out calculations of $\langle \Delta X(t)\rangle$ at room temperature for $n$$=$$10^9$ iterations using 
$X_0$$=$$0$ as initial condition, which corresponds to the initial molecular extension of $\Delta$$X(0)$$=$$0$,
$K_{sp}$$=$$40 pN/nm$, which is within the $20$$-$$50pN/nm$ range of values observed in the experimental force-extension 
curves of proteins. We used the time steps of $h$$=$$1$, $25$ and $50 ps$. We set $\kappa$$=$$10pN/nm$ and 
$\nu_0$$=$$1\mu m/s$, which translates to $r_f$$=$$10^{-5} pN/ns$. These are typical values of a cantilever spring 
constant and a pulling speed used in AFM experiments. We set the diffusion constant to 
$D$$=$$k_BT$$/\xi$$=$$1.5$$\times$$10^{-11}cm^2/s$, which corresponds to the force-driven $\sim$$150 nm$ extension 
of the fibrinogen molecule observed in AFM experiments over time $t$$=$$0.1 s$ \cite{WeiselBJ07}. The slow force-driven 
``diffusion of the molecular extension'', $\Delta$$X(t)$, described by the Brownian particle model with 
$D$$\approx$$10^{-11}cm^2/s$, should not be confused with the free Brownian diffusion of protein molecules in aqueous 
solution, for which $D$ is in the $10^{-6}$$-$$10^{-8}cm^2/s$ range.

We found that the average extensions $\langle \Delta
X(t_{n+1})\rangle_{sim}$, calculated using the first order
(Eq.~(\ref{eq:1st})) and the second order (Eq.~(\ref{eq:2nd}))
integrators, agree very well with each other and with the theoretical
curve of this quantity $\langle \Delta X(t)\rangle_{th}$
(Eq.~(\ref{eq:Xav})) for all values of $h$. The simulated data points
of $\langle \Delta X(t_{n+1})\rangle_{sim}$ practically collapsed on
the theoretical curve $\langle \Delta X(t)\rangle_{th}$ in the entire
range of time (data not shown). To quantify any reduction in accuracy,
we estimated the relative error of the molecular extension, $|\langle
\Delta X(t_{n+1})\rangle_{sim}$$-$$\langle \Delta
X(t)\rangle_{th}|/\langle \Delta X(t)\rangle_{th}$, accumulated at the
end of each trajectory and averaged over $10^5$ runs. The relative
error was found to be less than $2$$\times$$10^{-5}$
($1$$\times$$10^{-5}$) for the first-order (second-order) scheme,
significantly below the $10^{-3}$ level considered the acceptable
maximum for relative error in biomolecular simulations. These results
show that in a stochastic thermostat (random force) the numerical
integration errors, associated with the calculation of the global
mechanical reaction coordinate $\Delta X(t)$, are
minimal and/or cancel out, and that single precision arithmetic is
adequate for production runs. Hence, in the context of long
simulations of biomolecules on a GPU, the first-order integrator
(Ermak-McCammon algorithm) can be used to describe accurately their
mechanical properties under physiologically relevant conditions of
force application.


\subsection{Performance measurements}

We have compared the overall performance of an end-to-end application
of the SOP-GPU program with the heavily tuned CPU-based implementation
of the SOP model (SOP-CPU program) in describing the Langevin dynamics
of the domain $WW$ at equilibrium (Fig.~4, Table I). To fully occupy
the GPU resources, we profiled the computational performance of the
SOP-GPU program as a function of the number of independent
trajectories running concurrently on a single GPU. We refer to this as the
many-runs-per-GPU approach. Alternatively, we could have assessed the
performance of the program by running one trajectory on a single GPU,
but for a range of systems of different size $N$, which we refer to as
the one-run-per-GPU approach.  The results obtained indicate that for
a small system of $34$ residues ($WW$ domain, Table I), the use of a
single GPU device allows one to accelerate simulations starting from
$3$ independent runs (for small systems, there is insufficient
parallelism to fully load the GPU), which is also equivalent to
running one trajectory on a single GPU for a system of $\sim$$10^2$
residues, such as the domains $Ig27$ and $C2A$ (Table I). While the
simulation time on the CPU scales linearly with the number of runs,
the scaling in this regime on the GPU is sublinear (nearly constant)
until the number of runs is $\sim$$100$. At this point, depending on
the number of threads per thread block, the GPU shows significant
performance gains relative to the CPU reaching its maximum
$25$$-$$30$-fold value (the speedup is shown in the inset of
Fig.~4). We ran the simulations long enough to converge the speedup
ratio ($n$$=$$10^6$ steps of size $h$$=$$40 ps$). Beyond this point,
the GPU device is fully subscribed and the execution time scales
linearly with the number of runs as on the CPU.

In general, the total number of threads of execution $M_{th}$$=$$m_B$$B$, defined by the number of thread blocks 
$m_B$ of size $B$, is roughly equal the product of the system size $N$ and the number of trajectories $s$ running 
concurrently on the GPU, i.e. $M_{th}$$=$$N$$s$. Because it is impossible to predict which block size $B$ will 
result in the best performance, we carried out benchmark computations for $B$$=$$16$, $64$, and $256$ for purposes
of performance comparison. Our results indicate that all ALUs must to be fully loaded and that $M_{th}$ should 
exceed the number of ALUs on a single GPU by a factor of $10$$-$$15$. For example, on a graphics card with $240$ 
ALUs (GeForce GTX 280 or GTX 295, or Tesla C1060), $M_{th}$$\approx$$3,000$$-$$3,500$, and in the case of $WW$ domain 
($N$$=$$34$) this translates to $s$$\approx$$100$ (Fig.~4). This implies that the use of small thread blocks is 
more advantageous when $M_{th}$$\le$$3,000$, e.g., when simulating one trajectory for a system of the size of fibrinogen 
monomer $Fb$ ($N$$\approx$$2,000$) or $30$ runs for a system with $N$$\approx$$100$ ($Ig27$), or $\sim$$50$ runs 
for a smaller system such as the $WW$ domain ($N$$=$$34$). However, larger thread blocks should be used when 
$M_{th}$$>$$3,000$ to simulate a few trajectories for larger systems such as the fibrinogen dimer $(Fb)_2$ 
($N$$=$$3,849$) or to obtain one trajectory for a very large system, e.g., the viral capsid $HK97$ ($N$$=$$115,140$, 
see Table I).

To profile the associated computational time and memory demand for the SOP-GPU software we ran a series of benchmark 
simulations on several different protein systems (Table I) using the one-run-per-GPU approach. The execution time
of an end-to-end application of the program as a function of particle count was recorded using the standard CUDA 
runtime profiling tool. To more closely assess the performance characteristics of the SOP-GPU code as a function 
of the system size $N$, we analyzed one simulation run for each protein system, generated over $n$$=$$10^6$ steps of 
size $h$$=$$40 ps$ and block size of $B$$=$$64$ threads (Fig.~5). For all test systems of less than $\sim$$3,000$ 
residues the associated simulation time remains roughly the same, i.e. $(1-3)$$\times$$10^{-4}$ seconds per step. 
This is not surprising since $M_{th}$$\approx$$3,000$ is the amount of threads needed to fully utilize the GPU 
resources. For larger systems ($N$$>$$3,000$), the computational time scales roughly linearly with $N$ (Fig.~5). Small 
variation in runtime versus $N$ around the monotonic linear dependence is due to the different native topology of 
the test systems, i.e. the number of native and non-native contacts in the PDB structure (Table I). The amount of 
on-board memory in contemporary graphics cards - $\sim$$1 GB$ (GeForce GTX 200 series) and $4 GB$ (Tesla C1060) - 
is sufficient for Langevin simulations of large biomolecular systems comparable in size with the fibrinogen dimer 
$(Fb)_2$. The amount of on-board memory will most likely increase with the next generation of graphics cards with 
new Fermi architecture from NVIDIA (up to $\sim$$6 GB$) \cite{Fermi}.


\section{Conclusion}

We have developed and tested, to the best of our knowledge, the first GPU-based implementation of Langevin simulations
of biomolecules, where the particle based and the interacting pair based approaches have been employed in the parallel 
computation of forces due to the covalent and non-covalent interactions governed by the standard pair potentials. 
We have presented the numerical routines for the generation of pseudorandom numbers using the Hybrid Taus algorithm to
describe random collisions of a biomolecule with solvent molecules, for the construction of Verlet lists, and for the 
numerical integration of Langevin equations of motion based on the first order scheme. Although we focused on the 
$C_{\alpha}$-based coarse-grained SOP model of a protein, which involves only two-body potentials in the potential
energy function, the developed formalism can be used in conjunction with more sophisticated biomolecular force 
fields to explore, e.g., protein-protein and protein-DNA interactions, and it can also be extended to include the
three-body (angle) potentials to describe side chains. In addition, the numerical integration kernel can be 
modified to follow the Langevin dynamics in the underdamped limit, in order to describe the thermodynamics of
biomolecular transitions.

The developed formalism has been mapped into a standard CUDA code for Langevin simulations of biomolecules (SOP-GPU 
program). Benchmark simulations have shown that for a test system of the all-$\beta$ $WW$ domain the results of 
simulations of the mechanical denaturation on the GPU agree well with the results obtained on the CPU (Fig.~3). In a 
separate work, we also compared the forced unfolding data, obtained on the CPU and on the GPU, for the human synaptotagmin 
$Syt1$ (manuscript under review) and for the human fibrinogen $Fb$ molecules (manuscript in preparation), and found 
that the results of the CPU- and GPU-based computations agree very well. Using an exactly solvable 
model of a Brownian particle evolving in a harmonic potential, we have assessed the accuracy of the numerical 
integration of Langevin equations of motion in describing the force-induced elongation of a protein chain. We found
that using the first-order integrator is sufficient to accurately describe the force-driven elongation of a protein 
over many billions of iterations.

GPUs can be utilized to generate a few trajectories of Langevin dynamics for a large system of many thousands of 
residues (one-run-per-GPU approach) or many trajectories for a small system composed of a few hundreds of amino acids 
(many-runs-per-GPU approach). Using the one-run-per-GPU approach, we analyzed the relative CPU/GPU performance
of the program, and found that the GPU-based realization leads to a substantial $\sim$$25$$-$$30$ computational 
speedup, which depends on the number of threads in a thread block (similar acceleration can also be achieved using 
the one-run-per-GPU approach). We profiled the SOP-GPU program in terms of the computational time and the memory 
usage for a range of proteins (Table I). Our results show that the simulation time on a GPU remains roughly constant 
for a system of $N$$\approx$$10^2$$-$$10^3$ residues, and scales linearly with $N$ for a system of $N$$>$$10^3$ 
residues. The GPU on-board memory in contemporary graphics cards (GeForce 280, GeForce 295, and Tesla C1060) is 
sufficient to describe Langevin dynamics for large systems involving as many as $\sim$$10^5$ residues (Fig.~5). 
Describing mechanism(s) and multiple pathways underlying biomolecular transitions and resolving the entire distributions 
of the molecular characteristics requires gathering of the statistically significant amount of data. This can be 
achieved on a single GPU using the many-runs-per-GPU approach for smaller systems, and on multiple GPUs employing
the one-run-per-GPU approach for larger systems. The many-runs-per-GPU approach can also be utilized in conjunction 
with parallel tempering algorithms, including variants of the Replica Exchange Method, to resolve the phase diagrams 
of biomolecules.

The results obtained attest to the accuracy of the SOP-GPU program. The SOP-GPU software can now be utilized to
describe the mechanical properties of proteins, the strength of the noncovalent bonds that stabilize protein-protein 
complexes and aggregates, and the physical properties of large-size protein assemblies. A combination of the SOP model 
and the GPU-based computations enables one to carry out molecular simulations in reasonable wall-clock time in
order to explore the unfolding micromechanics of protein fibers and visco-elastic properties of biomolecular 
assemblies using the experimental protocol of force application. This allows one to interpret the experimental 
force-extension curves and force-indentation profiles of biomolecules, obtained in dynamic force spectroscopy
assays, thus, bridging the gap between theory and experiments. For example, on a GPU GeForce GTX 280 or GTX 295 
it takes only $\sim$$7$ days to generate a single unfolding trajectory for the fibrinogen monomer $Fb$ using 
experimental pulling speed of $2.5 \mu m/s$. By contrast, it would take as long as $\sim$$8$ months to obtain just
one trajectory using the CPU version of the program on a $2.66 GHz$ Intel Core i7 with $6 GB$ of memory. It takes 
$\sim$$35$ days to generate one force-indentation curve for the viral capsid $HK97$ on a single GPU (GeForce 
GTX $200$ series, Tesla C1060) using experimental pulling speed of $10 \mu m/s$. Hence, many force-indentation
trajectories can be generated on a single desktop computer equiped with several GPUs.

Beyond that, we note that due to rapid evolution of GPU hardware, the simulation time on the GPU will decrease 
significantly with the introduction of the MIMD (Multiple Instruction Multiple Data) based Fermi architecture (NVIDIA) 
in the near future \cite{Fermi}. The CUDA program, tested in this work, should be able to run with minor modifications
on future hardware making use of increased processing resources, and will permit the theoretical exploration of 
a range of interesting problems for which the experimental data are already available. The presented formalism 
can also be applied to the studies of other systems, including molecular motors, colloidosomes 
and liposomes.


\noindent {\bf Acknowledgments:} Acknowledgement is made to the donors of the American Chemical Society Petroleum 
Research Fund (grant PRF $\#$$47624$$-$$G6$) for partial support of this research (to VB). This project was also 
supported in part by the grant ($\#$$09$$-$$07$$12132$) from the Russian Foundation for Basic Research (to VB, YK, 
and AZ), and by the National Science Foundation grant MCB-0845002 (to RID).


\bibliographystyle{ieeetr}
\bibliography{papers,books}


\newpage

\section*{\bf FIGURE CAPTIONS}

\noindent
{\bf Fig.~1.}  Device architecture of a CPU (left panel) and a GPU (right panel). On a CPU, a significant 
amount of the chip surface provides cache and flow control for all ALUs. A CPU is capable of storing large 
amount of data on the DRAM (Dynamic Random Access Memory). Most of the GPU chip surface is devoted to
computational units, and a graphics card has on-board global memory. On the GPU, ALUs are grouped into 
multiprocessors, each of which has its own small flow control and cache units. To execute a simulation protocol 
on the GPU device, the CPU and the GPU communicate via the PCI Express bus, which allows the CPU to execute 
computational kernels on the GPU and to access GPU global memory.

\bigskip

\noindent
{\bf Fig.~2.}  The numerical algorithm and computational procedures
for the Langevin simulations of biomolecules on a GPU. The
computational workflow is shown using black arrows, and data transfer
- read and write operations from and to the DRAM and the HDD (hard
drive) on the CPU device and the GPU global memory - are represented
by the dashed arrows. Designing CUDA kernels involves the
decomposition of work into small fragments that can be mapped into
thread blocks, and further decomposition into warps and into
independent threads of execution. The computational workflow for only
one ($i$-th) thread running on the GPU and the workload division
between CPU and GPU are shown in detail. The execution of the program
is initiated on the CPU, which is used to prepare and store the
initial and output data. The CPU device starts the launch of each
computational kernel on the GPU for the calculation of forces,
generation of the random forces and Verlet lists, and for the
numerical integration of the Langevin equations of motion, using many
independent threads running in parallel.

\bigskip

\noindent
{\bf Fig.~3.}  Comparison of the results of pulling simulations for the all $\beta$ $WW$-domain (Table I) obtained 
on a CPU and on a GPU (the color code is explained in the graphs). Panel $(a)$: Representative examples of the 
dependence of mechanical tension experienced by the protein chain $f$ as a function of the molecular extension $R$ 
(force-extension curves) obtained using running averages over $500$ data points. Panel $(b)$: The histogram based
estimates of the distribution of unfolding forces $p(f^*)$, i.e., peak forces $f^*$ extracted from the force-extension 
curves. The histograms have been constructed using the bandwidth (bin size) of $h_{f^*}$$\approx$$3.6 pN$. Panel 
$(c)$: Representative examples of the time dependence of the average temperature of the protein chain $\langle T(t)\rangle$
(in units of $k_B$$T$), which correspond to the force-extension curves (panel $(a)$), obtained using running 
averages over $500$ data points.

\bigskip

\noindent
{\bf Fig.~4.}  The many-runs-per-GPU approach: shown are the simulation time on the CPU and on the GPU on a log-log 
scale and the relative CPU/GPU performance (computational speedup) of the end-to-end application of the SOP-GPU program 
(the inset) as a function of the number of equilibrium simulation runs for the all-$\beta$-strand 
$WW$ domain. While a single CPU core generates only one trajectory at a time, a GPU device is capable of
running many independent trajectories at the same time. The relative CPU/GPU performance is tested for the thread
block size of $B$$=$$16$, $64$, and $256$ computational threads of execution.

\bigskip

\noindent
{\bf Fig.~5.}  The one-run-per-GPU approach: shown are the simulation time on the GPU on a log-log scale and the memory 
usage (the inset) on a GPU as a function of the system size (number of residues) $N$. Results are
presented for several test systems, including small proteins (the $WW$-domain, the $Ig27$ domain from human titin, the 
domain $C2A$ from human synaptotagmin $Syt1$), large proteins fragments (the $\gamma$$C$ chain and the double-$D$ fragment 
from human fibrinogen $Fb$), long protein fibers (the $Fb$ monomer and dimer), and large-size protein assembly (the 
viral capsid $HK97$). The information about the native state topology for these biomolecules is summarized in Table I.


\newpage

\section*{}

\begin{table}
\parbox[t]{6.5in}
{\caption{Number of residues, covalent bonds, native contacts stabilizing the folded state, and 
residue pairs for a range of proteins ($WW$-domain, $Ig27$, $C2A$-domain, $\gamma$$C$ and $\beta$$C$ 
chains), protein fibers ($Fb$ monomer and dimer) and protein assembly (viral capsid $HK97$) used 
in the benchmark simulations.}}

\vspace{.2in}
\begin{tabular}{| c | c | c | c | c | c | c | c | c |}\hline
 Protein & $WW$\textsuperscript{\emph{a}} & $Ig27$\textsuperscript{\emph{b}} & $C2A$\textsuperscript{\emph{c}} & $\gamma$$C$\textsuperscript{\emph{d}} & $D$$-$$D$\textsuperscript{\emph{e}} & $Fb$\textsuperscript{\emph{f}} & $(Fb)_2$\textsuperscript{\emph{g}} &  $HK97$ \textsuperscript{\emph{h}} \\ \hline
 PDB code & 1PIN & 1TIT & 2R83\textsuperscript{\emph{i}} & 1M1J\textsuperscript{\emph{j}} & 1FZB & 3GHG & 3GHG & 1FT1\textsuperscript{\emph{k}} \\ \hline
 Residues & 34 & 89 & 126 & 517 & 1,062 & 1,913 & 3,849 & 115,140 \\ \hline
 Covalent bonds & 33 & 88 & 125 & 521 & 1,072 & 1,932 & 3,839 & 114,720 \\ \hline
 Native contacts &  65 & 255 & 328 & 1,770 & 3,498 & 5,709 & 12,560 & 467,904 \\ \hline
 Non-native pairs & 463 & 3,573 & 7,422 & 131,101 & 558,833 & 1,821,212 & 7,389,077 & 16,178,028\textsuperscript{\emph{l}} \\ \hline
\end{tabular}
\newline
\flushleft{
\textsuperscript{\emph{a}}{All-$\beta$-strand $WW$-domain.}\\
\textsuperscript{\emph{b}}{$Ig27$ domain of human titin.}\\
\textsuperscript{\emph{c}}{$C2A$-domain from human synaptotagmin $Syt1$.}\\
\textsuperscript{\emph{d}}{$\gamma$$C$ and $\beta$$C$ domains from human fibrinogen $Fb$.}\\
\textsuperscript{\emph{e}}{Double-$D$ fragment ($D$$-$$D$ interface) of human fibrinogen $Fb$.}\\
\textsuperscript{\emph{f}}{Human fibrinogen monomer $Fb$.}\\
\textsuperscript{\emph{g}}{Human fibrinogen dimer $(Fb)_2$ created from two $Fb$ monomers (3GHG) and the $D$$-$$D$ interface (1FZB).}\\
\textsuperscript{\emph{h}}{$HK97$ is Head II viral capsid.}\\
\textsuperscript{\emph{i}}{$C2A$ domain of human $Syt1$ protein.}\\
\textsuperscript{\emph{j}}{$\gamma$$C$ and $\beta$$C$ chains in human $Fb$ starting fom the $CYS$ ring in the $D$-domain.}\\
\textsuperscript{\emph{k}}{PDB code for a structural unit; the full $HK97$ capsid structure can be found in the Viper\cite{Viperdatabase} database.}\\
\textsuperscript{\emph{l}}{Based on a cut-off distance of $200$\AA.}\\
}

\end{table}


\newpage
\mbox{} 

\begin{figure}
\includegraphics[width=6.8in]{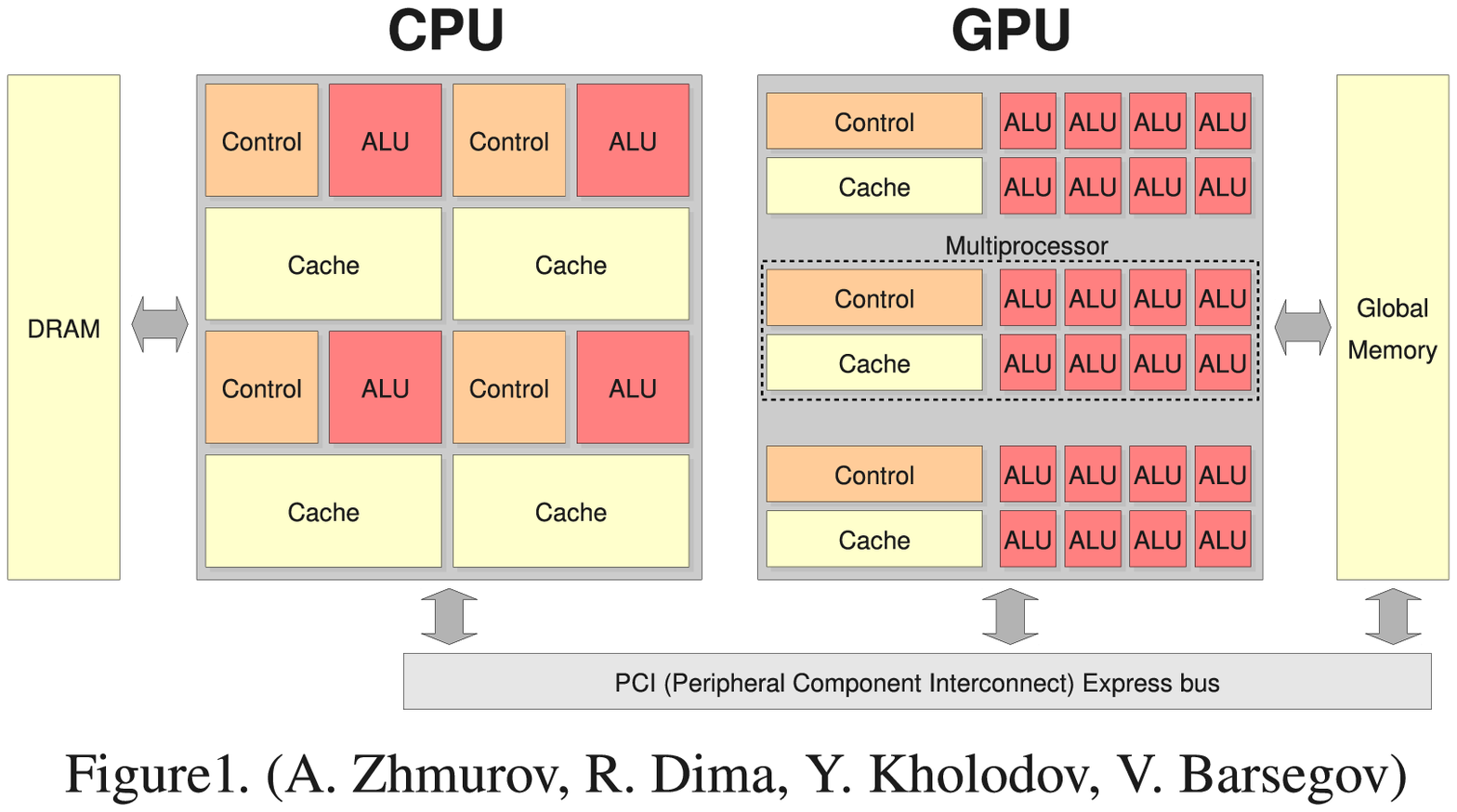}
\end{figure}

\newpage
\mbox{} 

\begin{figure}
\includegraphics[width=6.8in]{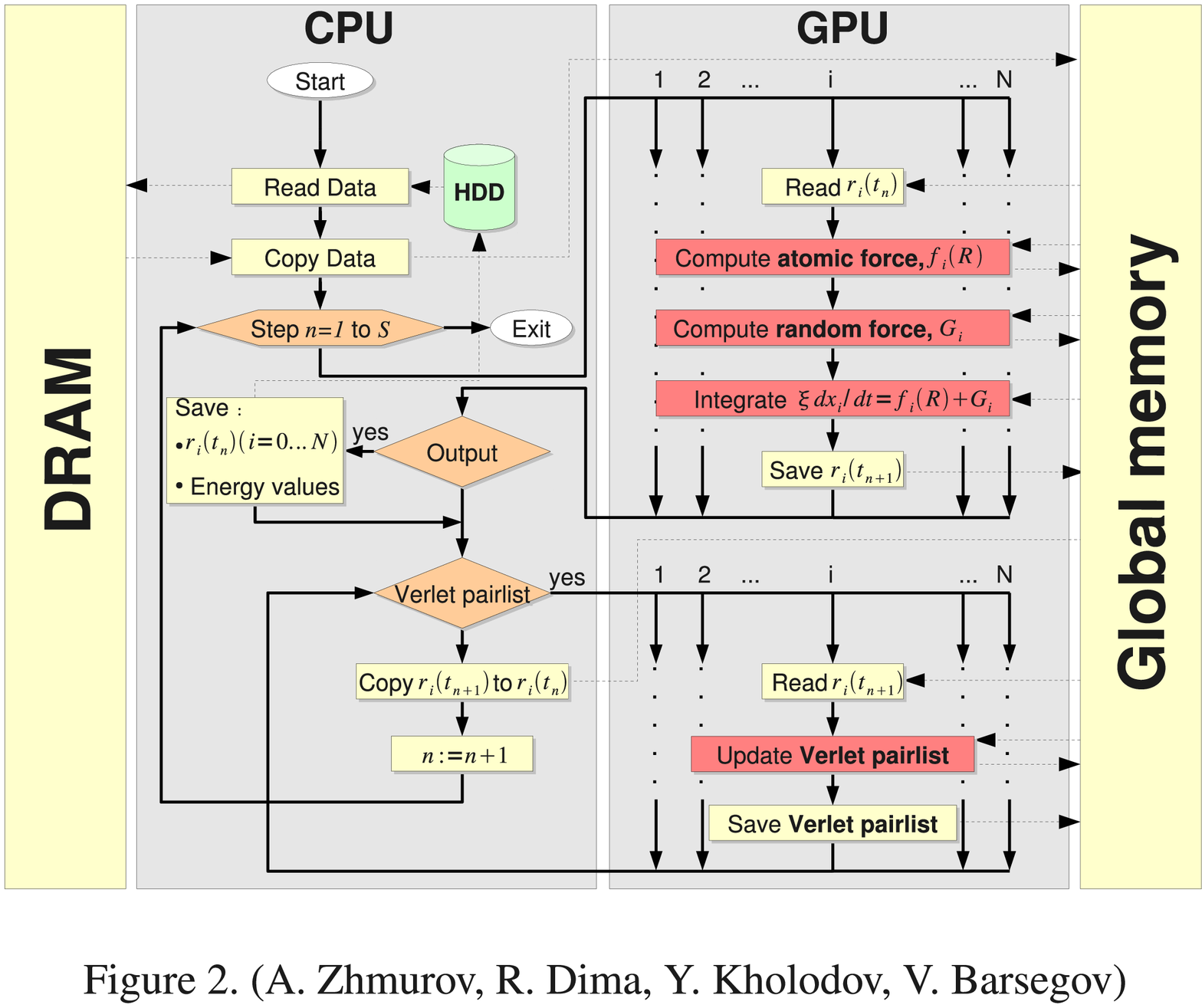}
\end{figure}


\begin{figure}
\includegraphics[width=4.5in]{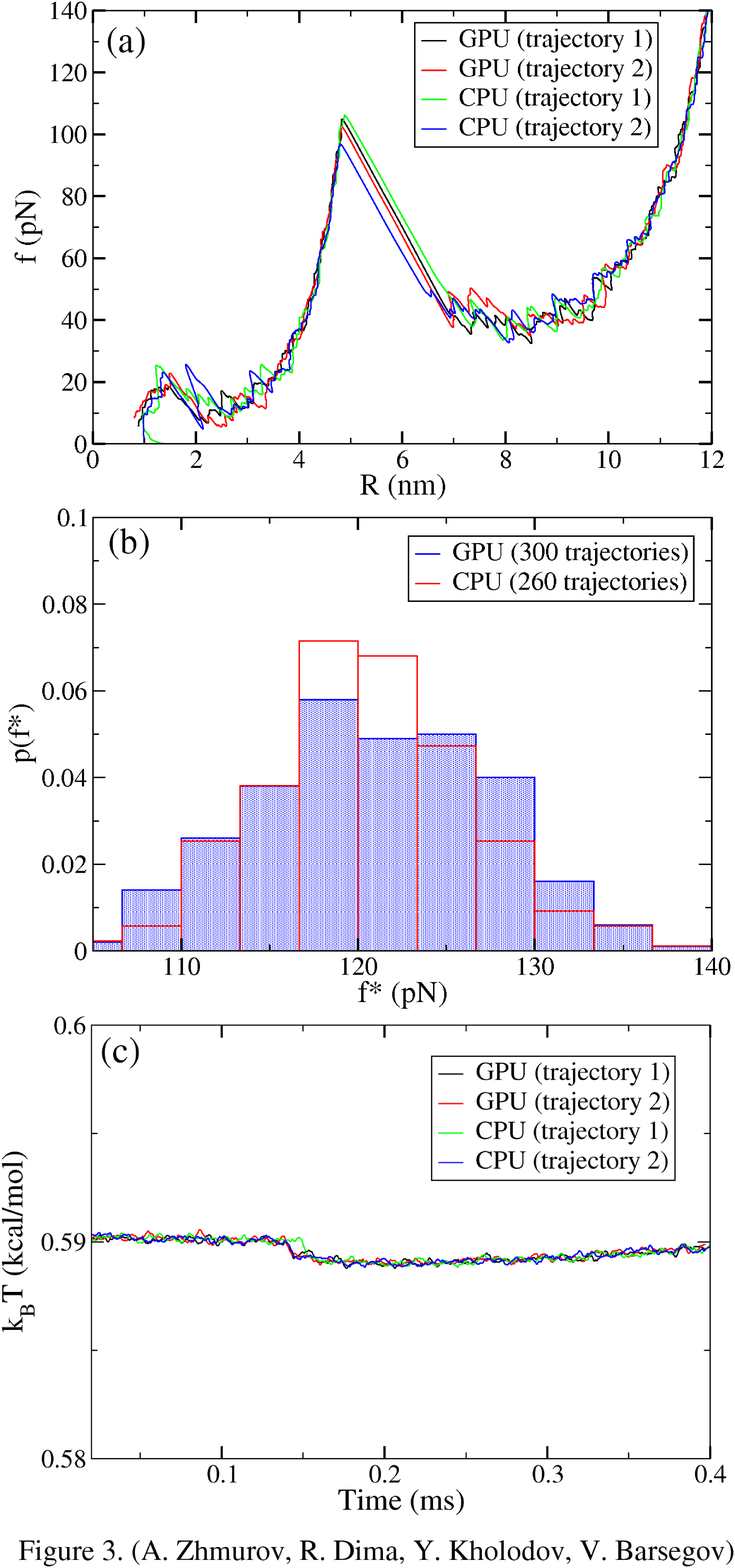}
\end{figure}

\newpage
\mbox{} 

\begin{figure}
\includegraphics[width=6.8in]{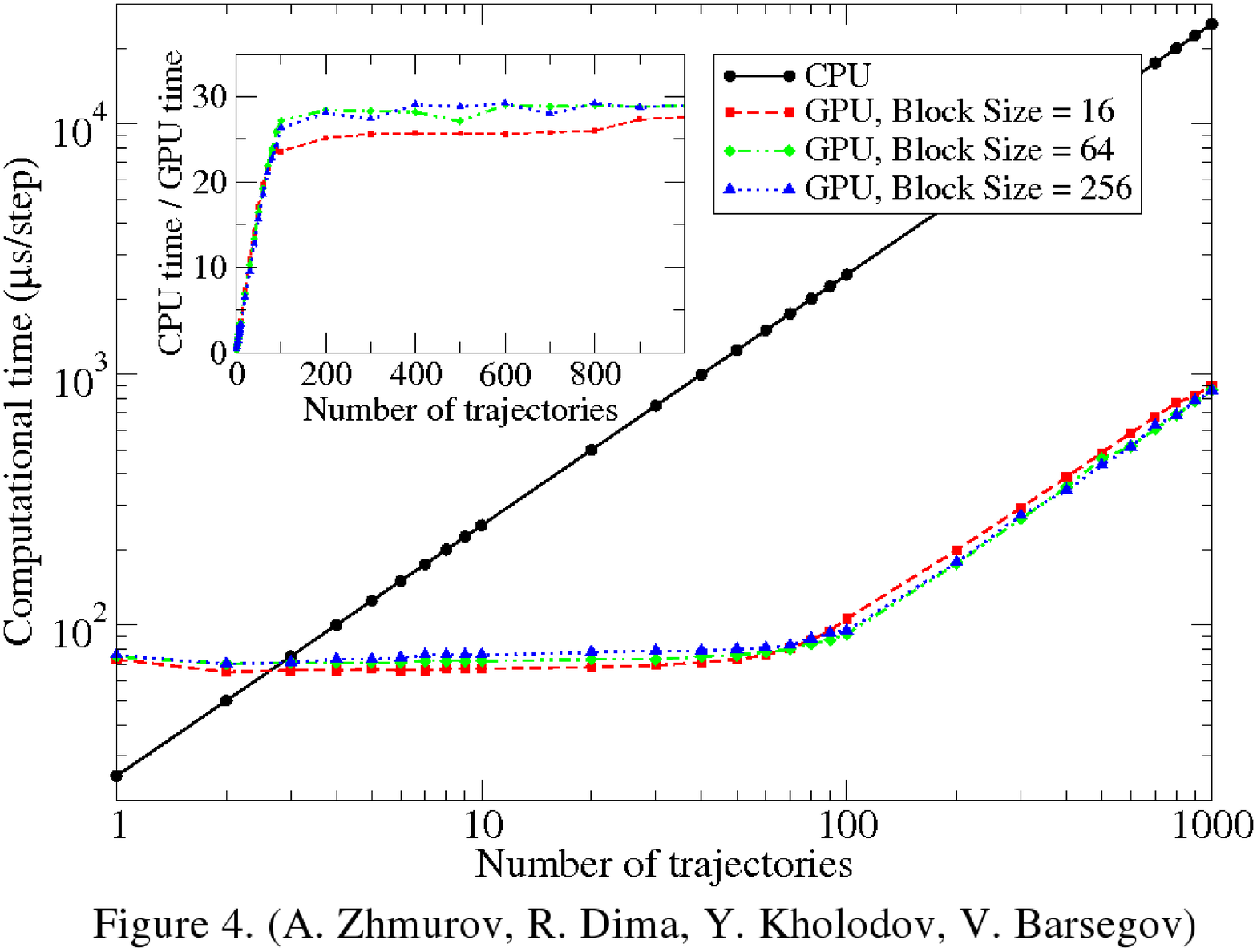}
\end{figure}

\newpage
\mbox{} 

\begin{figure}
\includegraphics[width=6.9in]{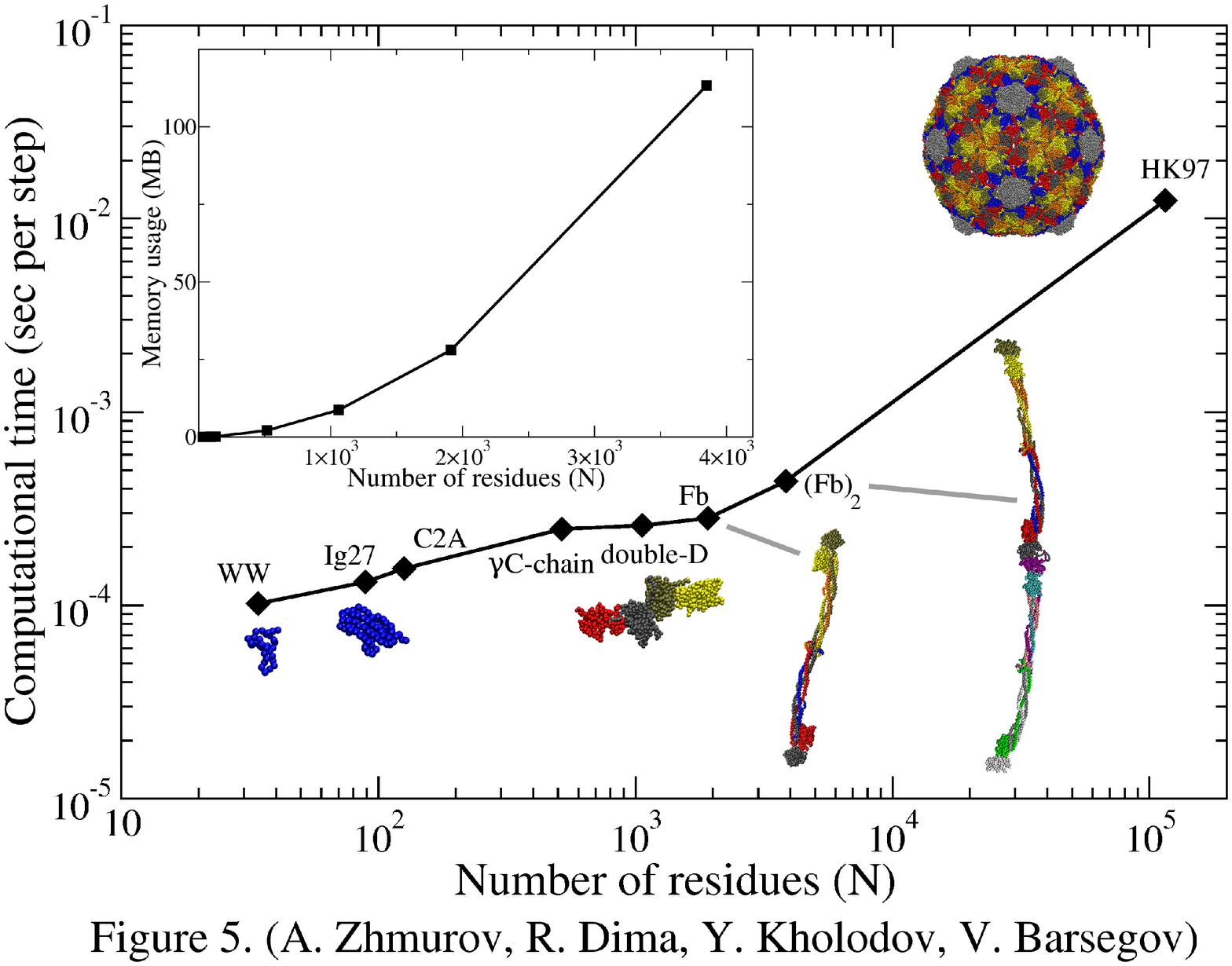}
\end{figure}


\end{document}